\documentclass{amsart}

\usepackage{amsfonts,amsmath,amssymb}
\usepackage{multirow}
\usepackage[longnamesfirst,round]{natbib}
\usepackage{graphicx}
\usepackage{lineno,hyperref}
\modulolinenumbers[5]
\usepackage{mathtools}
\usepackage{diagbox}
\usepackage{setspace}

\newcommand{\nrm}{\mathcal{N}}

\newcommand{\unif}{\mathcal{U}}

\newcommand{\vr}{\mathbb{V}}

\newcommand{\bs}{\boldsymbol}
\newcommand{\mbf}{\mathbf}

\newcommand{\diag}{\mathop{\mathrm{diag}}}

\newcommand{\bZ}{\bs{Z}}
\newcommand{\bU}{\bs{U}}

\newcommand{\bzero}{\bs{0}}

\newcommand{\ma}{\mbf{A}}

\newcommand{\momega}{\mbf{\Omega}}

\newcommand{\sed}{\text{\textsc{sed}}}

\newcommand{\bull}{\text{\raisebox{1pt}{\scalebox{.6}{$\bullet$}}}}

\title[R Package \texttt{krippendorffsalpha}]{\texttt{krippendorffsalpha}: An R Package for Measuring Agreement Using Krippendorff's Alpha Coefficient}

\author{John Hughes\\
Department of Statistics\\The Pennsylvania State University}
\email{drjphughesjr@gmail.com}

\begin{document}

\maketitle

\begin{abstract}
R package \texttt{krippendorffsalpha} provides tools for measuring agreement using Krippendorff's $\alpha$ coefficient, a well-known nonparametric measure of agreement (also called inter-rater reliability and various other names). This article first develops Krippendorff's $\alpha$ in a natural way, and situates $\alpha$ among statistical procedures. Then the usage of package \texttt{krippendorffsalpha} is illustrated via analyses of two datasets, the latter of which was collected during an imaging study of hip cartilage. The package permits users to apply the $\alpha$ methodology using built-in distance functions for the nominal, ordinal, interval, or ratio levels of measurement. User-defined distance functions are also supported. The fitting function can accommodate any number of units, any number of coders, and missingness. Bootstrap inference is supported, and the bootstrap computation can be carried out in parallel.
\end{abstract}

\section{Introduction}
\label{intro}

Krippendorff's $\alpha$ \citep{hayes2007answering} is a well-known nonparametric measure of agreement (i.e., consistency of scoring among two or more raters for the same units of analysis \citep{gwet}). In R \citep{Ihak:Gent:r::1996}, Krippendorff's $\alpha$ can be applied using function \texttt{kripp.alpha} of package \texttt{irr} \citep{irr}, function \texttt{kripp.boot} of package \texttt{kripp.boot} \citep{kripp.boot}, function \texttt{krippalpha} of package \texttt{icr} \citep{irr}, and functions \texttt{krippen.alpha.raw} and \texttt{krippen.alpha.dist} of package \texttt{irrCAC} \citep{irrcac}. But these packages fail to provide a number of useful features. In this article we present package \texttt{krippendorffsalpha}, which improves upon the above mentioned packages in (at least) the following ways. Package \texttt{krippendorffsalpha}
\begin{itemize}
\item offers commonly used built-in distance functions for the nominal, ordinal, interval, and ratio levels of measurement, and also supports user-defined distance functions;
\item conforms to the R idiom by providing S3 methods \texttt{confint}, \texttt{influence}, \texttt{plot}, and \texttt{summary};
\item supports embarrassingly parallel bootstrap computation; and
\item supports verbose communication with the user, including the display of a progress bar during the production of the bootstrap sample.
\end{itemize}

The remainder of this article is organized as follows. In Section~\ref{situate} we locate Krippendorff's $\alpha$ among statistical procedures. In Section~\ref{genesis} we first develop a special case of Krippendorff's $\alpha$ (call it $\alpha_\sed$) in a well-known parametric setting, and then we present $\alpha$ in its most general (i.e., nonparametric) form. In Section~\ref{mrpp} we show that $\alpha$ is a type of multiresponse permutation procedure. In Section~\ref{sklarsomega} we generalize $\alpha_\sed$ in a fully parametric fashion, arriving at Sklar's $\omega$. In Section~\ref{inference} we describe our package's bootstrap inference for $\alpha$ and compare the performance of our procedure to that of two alternative approaches. In Section~\ref{robust} we briefly discuss robustness and influence. In Section~\ref{illustrations} we provide a thorough demonstration of \texttt{krippendorffsalpha}'s usage before concluding in Section~\ref{summary}.

\section{Situating Krippendorff's Alpha among statistical procedures}
\label{situate}

Since Krippendorff's $\alpha$ is defined in terms of discrepancies \citep{krippendorff2013}, at first glance one might conclude, erroneously, that $\alpha$ is a measure of {\em dis}-agreement and so answers the wrong question. In Sections~\ref{genesis}--\ref{sklarsomega} we will show, by examining Krippendorff's $\alpha$'s place among statistical procedures, that $\alpha$ is, in fact, a sensible measure of agreement. Also, establishing a context for $\alpha$ may help practitioners make educated decisions regarding $\alpha$'s use.

The UML class diagram \citep{fowler2004uml} shown below in Figure~\ref{fig:kripp} provides a conceptual roadmap for our development. Briefly, a special case of $\alpha$ (which we denote as Alpha(SED) or $\alpha_\sed$) arises naturally in the context of the one-way mixed-effects ANOVA model. Alpha(SED) can then be generalized in a nonparametric fashion to arrive at Krippendorff's $\alpha$ as it has been presented by \citeauthor{hayes2007answering} (see \citet{gwet2015} for a development of nonparametric $\alpha$ in terms of agreement rather than discrepancies); this nonparametric form of $\alpha$ is a (slightly modified) multiresponse permutation procedure. Alternatively, $\alpha_\sed$ can be generalized in a parametric fashion to arrive at Sklar's $\omega$, a Gaussian copula-based methodology for measuring agreement.

\begin{figure}[ht]
   \centering
   \includegraphics[scale=.35]{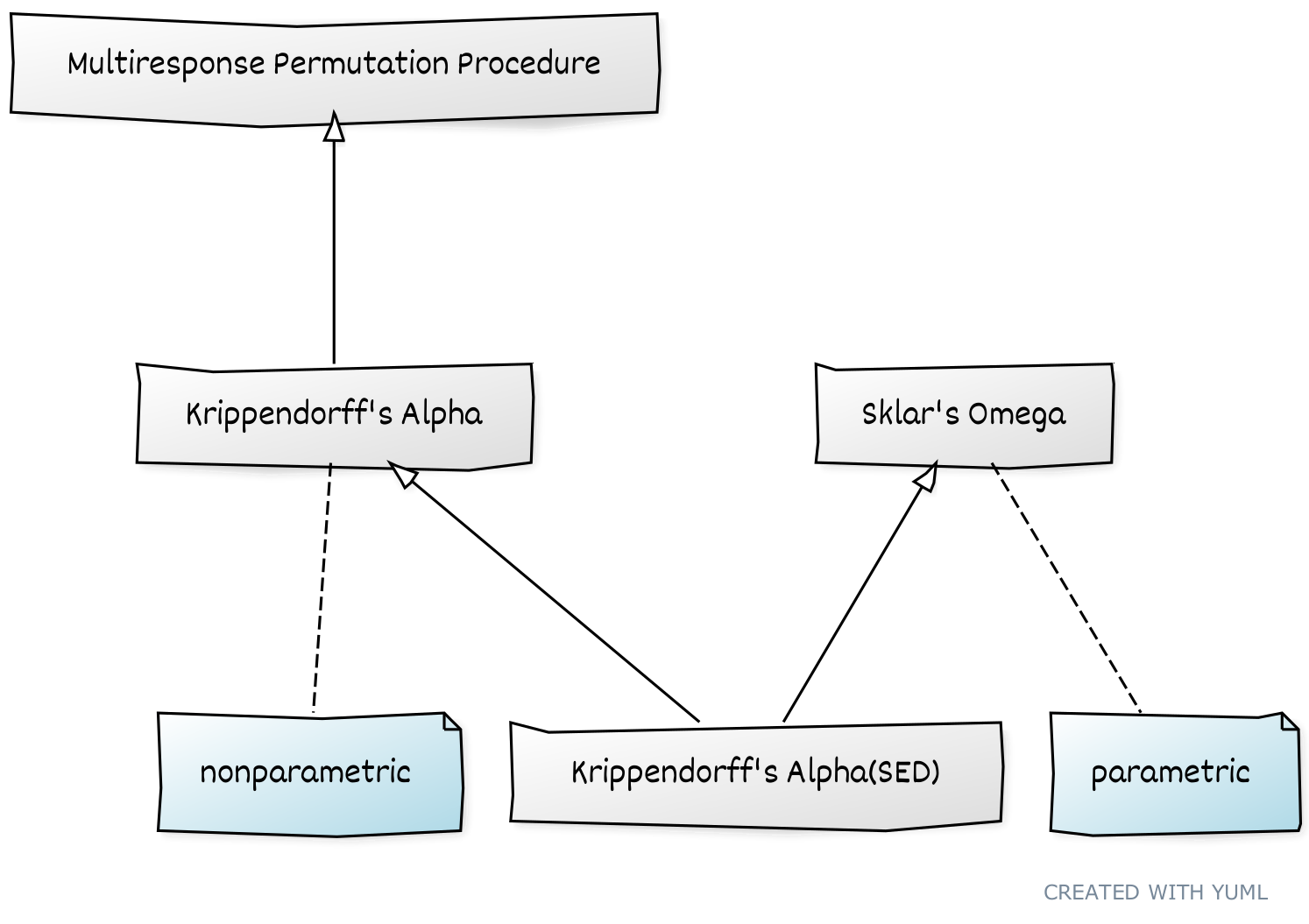}
   \caption{A UML class diagram that shows the relationships between Krippendorff's $\alpha$ and other statistical procedures.}
   \label{fig:kripp}
\end{figure}

\subsection{Parametric genesis of Krippendorff's Alpha coefficient}
\label{genesis}

In this section we develop Krippendorff's $\alpha$ \citep{hayes2007answering} in an intuitive and bottom-up fashion. Our starting point is a fully parametric model, namely, the classic one-way mixed-effects ANOVA model \citep{ravishanker2001first}. To ease exposition we will consider only a balanced version of the model. We have, for $n_u$ units and $n_c$ coders, scores
\[
Y_{ij} = \mu + \tau_i + \varepsilon_{ij},\;\;\;\;\;(i=1,\dots,n_u;\; j = 1,\dots,n_c)
\]
where
\begin{itemize}
\item $\mu$ (the population mean) is a fixed real number,
\item the $\tau_i$ are independent $\nrm(0,\sigma_\tau^2)$ random variables,
\item the $\varepsilon_{ij}$ are independent $\nrm(0,\sigma_\varepsilon^2)$ random variables, and
\item the $\tau_i$ are independent of the $\varepsilon_{ij}$.
\end{itemize}
In this setup we have $n_c$ Gaussian codes $Y_{i1},\dots,Y_{in_c}$ for unit $i\in\{1,\dots,n_u\}$. Conditional on $\tau_i$, said codes are $\nrm(\mu+\tau_i,\sigma_\varepsilon^2)$ random variables. Since the variables share the ``unit effect'' $\tau_i$, the variables are correlated. The correlation, which is usually called the {\em intraclass correlation}, is given by
\[
\alpha=\frac{\sigma_\tau^2}{\sigma_\tau^2+\sigma_\varepsilon^2}=1-\frac{\sigma_\varepsilon^2}{\sigma_\tau^2+\sigma_\varepsilon^2}.
\]
We use $\alpha$ to denote this quantity precisely because Krippendorff's $\alpha$ is the intraclass correlation for codes that conform to this model. That is, for the one-way mixed-effects ANOVA model, Krippendorff's $\alpha$ is the intraclass correlation. The reader may recall that the estimator of $\alpha$ for this model is
\begin{align}
\label{intraclass}
\hat{\alpha}&=1-\frac{\widehat{\sigma_\varepsilon^2}}{\widehat{\sigma_\tau^2+\sigma_\varepsilon^2}}=1-\frac{\frac{1}{n_u(n_c-1)}\sum_{i=1}^{n_u}\sum_{j=1}^{n_c}(Y_{ij}-\bar{Y}_{i\bull})^2}{\frac{1}{n_un_c-1}\sum_{i=1}^{n_u}\sum_{j=1}^{n_c}(Y_{ij}-\bar{Y}_{\bull\bull})^2},
\end{align}
where $\bar{Y}_{i\bull}$ and $\bar{Y}_{\bull\bull}$ denote the arithmetic means for the $i$th unit and for the entire sample, respectively. The form of this estimator is not surprising, of course, since it is well known that assuming Gaussianity leads to variance estimators involving sums of weighted squared deviations from sample arithmetic means.

We can eliminate the arithmetic means in (\ref{intraclass}) by employing the identity
\[
\sum_{i=1}^n(x_i-\bar{x}_\bull)^2=\frac{1}{2n}\sum_{i=1}^n\sum_{j=1}^n(x_i-x_j)^2.
\]
This gives
\begin{align}
\label{intranew}
\hat{\alpha}&=1-\frac{\frac{1}{2n_un_c(n_c-1)}\sum_{i=1}^{n_u}\sum_{j=1}^{n_c}\sum_{k=1}^{n_c}(Y_{ij}-Y_{ik})^2}{\frac{1}{2n_un_c(n_un_c-1)}\sum_{i=1}^{n_u}\sum_{j=1}^{n_c}\sum_{k=1}^{n_u}\sum_{l=1}^{n_c}(Y_{ij}-Y_{kl})^2}.
\end{align}
Now, let $d^2(x,y)=(x-y)^2$ and rewrite (\ref{intranew}) as
\begin{align}
\label{sed}
\hat{\alpha}&=1-\frac{D_o}{D_e}=1-\frac{\frac{1}{2n_un_c(n_c-1)}\sum_{i=1}^{n_u}\sum_{j=1}^{n_c}\sum_{k=1}^{n_c}d^2(Y_{ij},Y_{ik})}{\frac{1}{2n_un_c(n_un_c-1)}\sum_{i=1}^{n_u}\sum_{j=1}^{n_c}\sum_{k=1}^{n_u}\sum_{l=1}^{n_c}d^2(Y_{ij},Y_{kl})},
\end{align}
where $D_o$ and $D_e$ denote observed and expected disagreement, respectively. This is Krippendorff's $\alpha$ for the squared Euclidean distance (which is not a metric but a Bregman divergence \citep{bregman}). (We will henceforth refer to this version of $\alpha$ as Alpha(SED) or $\alpha_\sed$.) As we mentioned above, this form of Krippendorff's $\alpha$ arises quite naturally when the data at hand conform to the one-way mixed-effects ANOVA model, for which agreement corresponds to positive correlation. More generally, Krippendorff recommends this form of $\alpha$ for the interval level of measurement. For other levels of measurement, Krippendorff presents other distance functions $d^2$ (several possibilities are shown in Table~\ref{distance}). Note that package \texttt{krippendorffsalpha} supports user-defined distance functions as well as the interval, nominal, and ratio distance functions shown in the table.
\begin{table}[h]
   \centering
   \begin{tabular}{cl}
      Level of Measurement   & Distance Function \\\hline
     interval & $d^2(x,y)=(x-y)^2$\\
      nominal & $d^2(x,y)=1\{x\neq y\}$\\
      ratio & $d^2(x,y)=\left(\frac{x-y}{x+y}\right)^2$\\
      bipolar & $d^2(x,y)=\frac{(x-y)^2}{(x+y-2x_\text{min})(2x_\text{max}-x-y)}$\\
      circular & $d^2(x,y)=\left\{\sin\left(\pi\frac{x-y}{I}\right)\right\}^2\;\;(I=$ number of equal intervals on circle)\\
      ordinal & $d^2(x,y)=(x-y)^2\;\;$ (adjacent ranks are equidistant)
         \end{tabular}
   \caption{Several distance functions that may be appropriate for use in Krippendorff's $\alpha$.}
   \label{distance}
\end{table}

\subsection{Alpha as a multiresponse permutation procedure}
\label{mrpp}

In any case, (\ref{sed}) is nonparametric for arbitrary $d^2$ since then the estimator $\hat{\alpha}$ does not usually correspond to a well-defined population parameter $\alpha$.  This more general form of Krippendorff's $\alpha$ is in fact a special case of the so called multiresponse permutation procedure (MRPP). The MRPPs form a class of permutation methods for discerning differences among two or more groups in one or more dimensions \citep{permutation}. Note, however, that although $\alpha$ can be viewed as an MRPP (as we are about to show), $\alpha$ has been modified for the purpose of measuring agreement rather than discerning differences.

To show that Krippendorff's $\alpha$ is an MRPP, we first present the general form of the MRPP. Following \citeauthor{permutation}, let $\Omega=\{\omega_1,\dots,\omega_n\}$ be a finite sample that is representative of some population of interest, let $S_1,\dots,S_{a+1}$ denote a partition of $\Omega$ into $a+1$ disjoint groups, and let $\rho$ be a metric that makes sense for the objects of $\Omega$. (Strictly speaking, $\rho$ need not be a metric; a symmetric distance function will suffice.) To ease notation a bit, let $\rho_{jk}\equiv\rho(\omega_j,\omega_k)$. Then the MRPP statistic can be written as
\[
\delta=\sum_{i=1}^aC_i\theta_i,
\]
where $C_i>0$ are group weights such that $\sum_iC_i=1$;
\[
\theta_i=\frac{1}{\begin{pmatrix}n_i\\2\end{pmatrix}}\sum_{j<k}\rho_{jk}1_i\{\omega_j\}1_i\{\omega_k\},
\]
is the average distance between distinct pairs of objects in group $S_i$; $n_i\geq 2$ is the number of objects in group $i$; $l=\sum_{i=1}^an_i$; $n_{a+1}=n-l\geq 0$ is the number of remaining unclassified objects in group $S_{a+1}$; and $1_i$ is the indicator function for membership in group $i$.

Note that this formulation is quite general since the objects in $\Omega$ can be scalars, vectors, or more exotic objects; and we are free to choose the metric and the weights. In the case of Krippendorff's $\alpha$ we can produce $\delta=D_o$ by letting $\rho=d^2$ for some appropriately chosen distance function $d^2$, and choosing weights $C_i=1/n_u$.

\subsection{A parametric generalization of Alpha(SED)}
\label{sklarsomega}

In the preceding sections we generalized $\alpha_\sed$ in a nonparametric fashion by substituting other notions of distance for the squared Euclidean distance. Now we will present a fully parametric generalization of $\alpha_\sed$, namely, Sklar's $\omega$ \citep{sklarsomega}.

The statistical model underpinning Sklar's $\omega$ is a Gaussian copula model \citep{xue2000multivariate}. The most general form of the model is given by
\begin{align}
\label{gausscop}
\nonumber\bZ = (Z_1,\dots,Z_n)'  & \; \sim\;  \nrm(\bzero,\momega)\\
\nonumber U_i = \Phi(Z_i) & \;\sim\; \unif(0,1)\;\;\;\;\;\;\;(i=1,\dots,n)\\
Y_i = F_i^{-1}(U_i) & \;\sim\; F_i,
\end{align}
where $\momega$ is a correlation matrix, $\Phi$ is the standard Gaussian cdf, and $F_i$ is the cdf for the $i$th outcome $Y_i$. Note that $\bU=(U_1,\dots, U_n)'$ is a realization of the Gaussian copula, which is to say that the $U_i$ are marginally standard uniform and exhibit the Gaussian correlation structure defined by $\momega$. Since $U_i$ is standard uniform, applying the inverse probability integral transform to $U_i$ produces outcome $Y_i$ having the desired marginal distribution $F_i$.

To see that the one-way mixed-effects ANOVA model (and hence $\alpha_\sed$) is a special case of Sklar's $\omega$, let the copula correlation matrix  $\momega$ be block diagonal, where the $i$th block corresponds to the $i$th unit $(i=1,\dots,n_u)$ and has a compound symmetry  structure. That is,
\[
\momega = \diag(\momega_i),
\]
where
\[
\momega_i = \bordermatrix{ & c_1 & c_2 & \dots & c_{n_c} \cr
c_1 & 1 & \omega  &\dots & \omega\cr
c_2 & \omega &  1 &  \dots & \omega\cr
\vdots & \vdots & \vdots  & \ddots  & \vdots\cr
c_{n_c} & \omega &  \omega & \dots  & 1
}.
\]
Complete the specification by letting $F_{ij}\;(i=1,\dots,n_u;\;j=1,\dots,n_c)$ be the cdf for the Gaussian distribution with mean $\mu$ and variance $\sigma^2$. Then $\omega=\alpha$, the intraclass correlation coefficient.

\section{Inference for Krippendorff's Alpha}
\label{inference}

\citeauthor{permutation} describe hypothesis testing for MRPPs. Specifically, they discuss three approaches: permutation, Monte Carlo resampling, and Pearson type III moment approximation; the latter of which has significant advantages. For Krippendorff's $\alpha$, though, we are interested not in hypothesis testing but in interval estimation. This can be done straightforwardly and efficiently using Monte Carlo resampling. Since $D_e$ is invariant to permutation of the scores, our resampling procedure focuses on $D_o$ only. The algorithm proceeds as follows.
\begin{enumerate}
\item Collect the scores in an $n_u\times n_c$ matrix, $\ma$, where each row corresponds to a unit.
\item For $i\in\{1,\dots,n_b\}$, form matrix $\ma_i$ by sampling, with replacement, $n_u$ rows from $\ma$.
\item For each $\ma_i$, compute $D_o^{(i)}$ using the same distance function $d^2$ that was used to compute $\hat{\alpha}$.
\item For each $D_o^{(i)}$, compute $\hat{\alpha}_i=1-D_o^{(i)}/D_e$.
\end{enumerate}
The resulting collection $\{\hat{\alpha}_1,\dots,\hat{\alpha}_{n_b}\}$ is a bootstrap sample for $\hat{\alpha}$, sample quantiles of which are estimated confidence limits for $\alpha$.

We carried out a number of realistic simulation experiments and found that this approach to interval estimation performs well in a wide variety of circumstances. When the true distribution of $\hat{\alpha}$ is (at least approximately) symmetric, \citeauthor{gwet2015}'s closed-form expression for $\hat{\vr}(\hat{\alpha})$, which is implemented (for categorical data only) in package \texttt{irrCAC}, also performs well. By contrast, we found that the bootstrapping procedure recommended by \citet{alphaboot}, which is implemented in packages \texttt{kripp.boot} and \texttt{icr}, generally performs rather poorly, producing intervals that are far too narrow (e.g., 95\% intervals achieve 74\% coverage). 

\section{Robustness and interpretation}
\label{robust}

For some levels of measurement, one may,  in the interest of robustness, be tempted to replace squares with absolute values (in the distance function $d^2$). This would be advantageous if one aimed to do hypothesis testing. But for Krippendorff's $\alpha$, using absolute values instead of squares proves disastrous, for the resulting estimator $\hat{\alpha}$ is substantially negatively biased and tends to lead to erroneous inference regarding agreement. All is not lost, however, since package \texttt{krippendorffsalpha} provides a means of investigating the influence on $\hat{\alpha}$ of any unit or coder (see the next section for examples).

\section{Illustrations}
\label{illustrations}

Here we illustrate the use of \texttt{krippendorffsalpha} by applying Krippendorff's $\alpha$ to a couple of datasets. We will interpret the results according to the ranges given in Table~\ref{tab:interpret}, but we suggest---as do Krippendorff and others \citep{artstein2008inter,landiskoch}---that an appropriate reliability threshold may be context dependent.
\begin{table}[h]
\centering
\begin{tabular}{cl}
Range of Agreement & Interpretation\\\hline
$\phantom{0.2<\;}\alpha\leq 0.2$ & Slight Agreement\\
$0.2<\alpha\leq 0.4$ & Fair Agreement\\
$0.4<\alpha\leq 0.6$ & Moderate Agreement\\
$0.6<\alpha\leq 0.8$ & Substantial Agreement\\
$\phantom{0.2<\;}\alpha>0.8$ & Near-Perfect Agreement\\
\end{tabular}
\caption{Guidelines for interpreting values of an agreement coefficient.}
\label{tab:interpret}
\end{table}

\subsection{Nominal data analyzed previously by Krippendorff}

Consider the following data, which appear in \citep{krippendorff2013}. These are nominal values (in $\{1,\dots,5\}$) for twelve units and four coders. The dots represent missing values.

\begin{figure}[h]
   \centering
   \begin{tabular}{ccccccccccccc}
   & $u_1$ &  $u_2$ & $u_3$ & $u_4$ & $u_5$ & $u_6$ & $u_7$ & $u_8$ & $u_9$ & $u_{10}$ & $u_{11}$ & $u_{12}$\vspace{2ex}\\
   $c_1$ & 1 & 2 & 3 & 3 & 2 & 1 & 4 & 1 & 2 & \bull & \bull & \bull\\
   $c_2$ & 1 & 2 & 3 & 3 & 2 & 2 & 4 & 1 & 2 & 5 & \bull & 3\\
   $c_3$ & \bull & 3 & 3 & 3 & 2 & 3 & 4 & 2 & 2 & 5 & 1 & \bull\\
   $c_4$ & 1 & 2 & 3 & 3 & 2 & 4 & 4 & 1 & 2 & 5 & 1 & \bull
   \end{tabular}
   \caption{Some example nominal outcomes for twelve units and four coders, with seven missing values.}
   \label{fig:nominal}
\end{figure}

Note that the scores for all units except the sixth are constant or nearly so. This suggests near-perfect agreement, and so we should expect $\hat{\alpha}$ to be greater than 0.8.

To apply Krippendorff's $\alpha$ to these data, first we load package \texttt{krippendorffsalpha}.

\bigskip\begin{verbatim}
R> library(krippendorffsalpha)

krippendorffsalpha: Measuring Agreement Using Krippendorff's Alpha Coefficient
Version 1.1 created on 2021-01-13.
copyright (c) 2020-2021, John Hughes
For citation information, type citation("krippendorffsalpha").
Type help(package = krippendorffsalpha) to get started.
\end{verbatim}\bigskip

Now we create the dataset as a matrix such that each row corresponds to a unit and each column corresponds to a coder.

\bigskip\begin{verbatim}
R> nominal = matrix(c(1,2,3,3,2,1,4,1,2,NA,NA,NA,
+                     1,2,3,3,2,2,4,1,2,5,NA,3,
+                     NA,3,3,3,2,3,4,2,2,5,1,NA,
+                     1,2,3,3,2,4,4,1,2,5,1,NA), 12, 4)
R> nominal

      [,1] [,2] [,3] [,4]
 [1,]    1    1   NA    1
 [2,]    2    2    3    2
 [3,]    3    3    3    3
 [4,]    3    3    3    3
 [5,]    2    2    2    2
 [6,]    1    2    3    4
 [7,]    4    4    4    4
 [8,]    1    1    2    1
 [9,]    2    2    2    2
[10,]   NA    5    5    5
[11,]   NA   NA    1    1
[12,]   NA    3   NA   NA
\end{verbatim}\bigskip

Next we apply Krippendorff's $\alpha$ for the nominal level of measurement. If argument \texttt{level} is set to \texttt{"nominal"}, the discrete metric $d^2(x,y)=1\{x\neq y\}$ is used by default. We do a bootstrap with sample size $n_b=$ 1,000 (argument \texttt{confint} defaults to \texttt{TRUE}, and control parameter \texttt{bootit} defaults to 1,000). We set control parameter \texttt{parallel} equal to \texttt{FALSE} because the dataset is too small to warrant parallelization of the bootstrap computation. Finally, we set argument \texttt{verbose} equal to \texttt{TRUE} so that a progress bar is shown during the bootstrap computation. The computation took less than one second.

\bigskip\begin{verbatim}
R> set.seed(42)
R> fit.full = krippendorffs.alpha(nominal, level = "nominal",
+                                 control = list(parallel = FALSE),
+                                 verbose = TRUE)

  |++++++++++++++++++++++++++++++++++++++++++++++++++| 100% elapsed=00s
\end{verbatim}\bigskip

As is customary in R, one can view a summary by passing the fit object to \texttt{summary.krippendorffsalpha}, an S3 method. If \texttt{krippendorffs.alpha} was called with \texttt{confint = TRUE}, \texttt{summary} displays a 95\% confidence interval by default. The confidence level can be specified using argument \texttt{conf.level}. In any case, the quantile method \citep{davison1997bootstrap} is used to estimate the confidence limits. Any arguments passed to \texttt{summary.krippendorffsalpha} via $\dots$ are passed on to R's \texttt{quantile} function; this allows the user to control, for example, how the sample quantiles are computed.

\bigskip\begin{verbatim}
R> summary(fit.full)

Krippendorff's Alpha

Data: 12 units x 4 coders

Call:

krippendorffs.alpha(data = nominal, level = "nominal", verbose = TRUE, 
    control = list(parallel = FALSE))

Control parameters:
              
parallel FALSE
bootit   1000 
              
Results:

      Estimate  Lower Upper
alpha   0.7429 0.4644     1
\end{verbatim}\bigskip

We see that $\hat{\alpha}=0.74$ and $\alpha\in(0.46, 1.00)$. This point estimate indicates only substantial agreement, which is not what we expected. At least the interval is consistent with near-perfect agreement, but we should not take this interval too seriously since the interval is rather wide (owing to the small size of the dataset).

Perhaps the substantial disagreement for the sixth unit was influential enough to yield $\hat{\alpha}\leq 0.8$. We can use \texttt{influence.krippendorffsalpha}, another S3 method, to investigate. This function, like other R versions of \texttt{influence} (e.g., \texttt{influence.lm}, \texttt{influence.glm}), computes DFBETA statistics \citep{Young2017Handbook-of-Reg}, as illustrated below.

\bigskip\begin{verbatim}
R> (inf.6 = influence(fit.full, units = 6))

$dfbeta.units
         6 
-0.1141961 
\end{verbatim}\bigskip

Leaving out the sixth unit yields a DFBETA statistic of -0.11, which implies that $\hat{\alpha}$ would have been 0.86. This is consistent with our initial hypothesis.

\bigskip\begin{verbatim}
R> fit.full$alpha.hat - inf.6$dfbeta.units

    alpha 
0.8571429 
\end{verbatim}\bigskip

Let us call \texttt{krippendorffs.alpha} again to get a new interval.

\bigskip\begin{verbatim}
R> fit.sub = krippendorffs.alpha(nominal[-6, ], level = "nominal",
+                                control = list(parallel = FALSE))
confint(fit.sub)

    0.025     0.975 
0.6616541 1.0000000
\end{verbatim}\bigskip

We see that excluding the sixth unit leads to $\alpha\in(0.66,1.00)$. The new 95\% interval was returned by S3 method \texttt{confint.krippendorffsalpha}, whose \texttt{level} argument defaults to 0.95, in keeping with R's other \texttt{confint} methods. Note that \texttt{confint.krippendorffsalpha}, like \texttt{summary.krippendorffsalpha}, passes any $\dots$ arguments on to the \texttt{quantile} function.

We conclude this example by producing a visual display of our results (Figure~\ref{fig:nominal}). (The figure was produced via a call to S3 method \texttt{plot.krippendorffsalpha}, which in turn calls \texttt{hist} and \texttt{abline}, and does not show a kernel density estimate. Function \texttt{plot.krippendorffsalpha} is capable of producing highly customized plots; see the package documentation for details.) Since $\hat{\alpha}$ is close to 1 and the dataset is small, the bootstrap distribution is substantially skewed to the left. Thus these data provide a textbook example of the importance of bootstrapping.

\bigskip\begin{verbatim}
R> plot(fit.sub, xlim = c(0, 1), xlab = "Bootstrap Estimates", main = "Nominal Data",
+       density = FALSE)
\end{verbatim}\bigskip

\begin{figure}[ht]
   \centering
   \includegraphics[scale=.6]{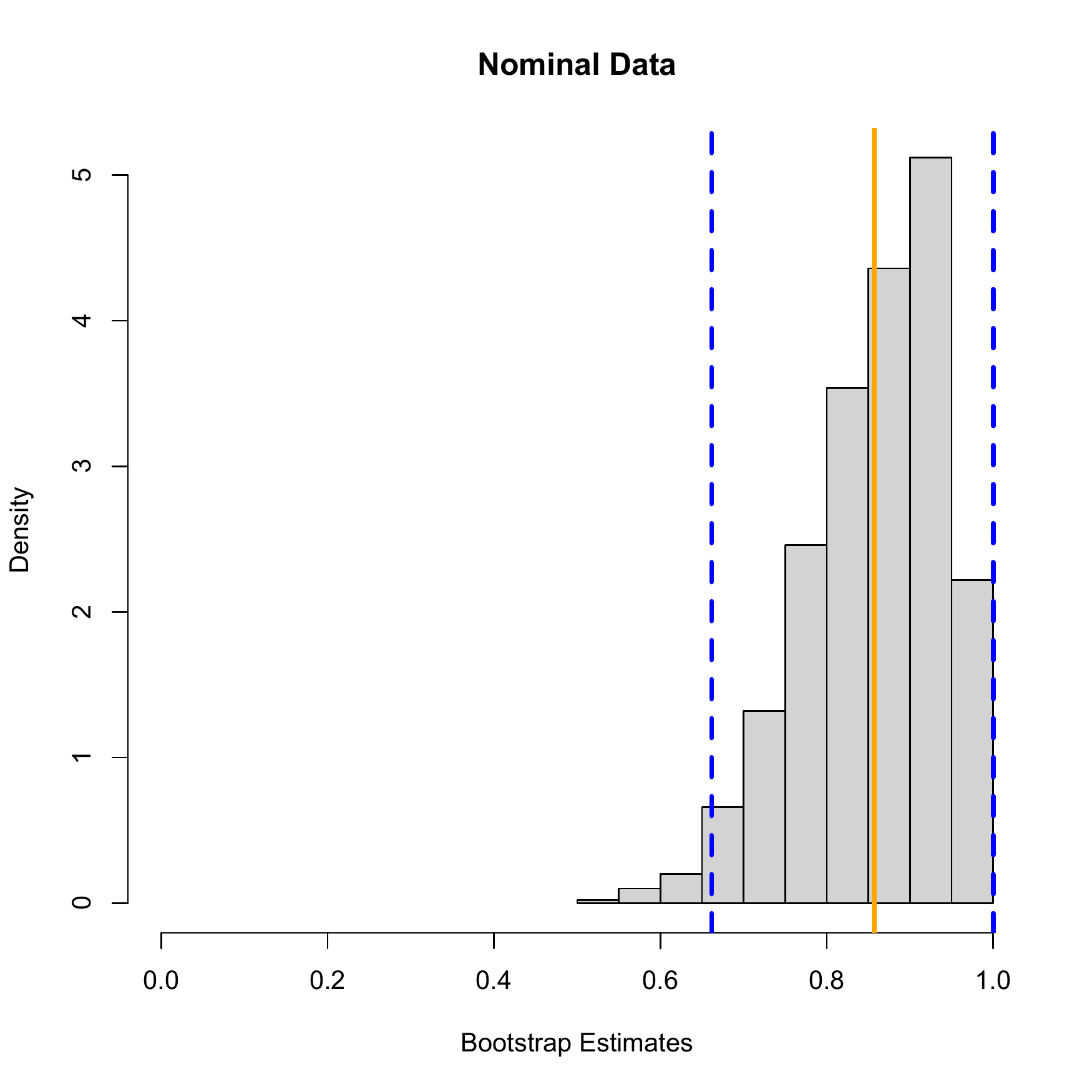}
   \caption{A plot of the results from our analysis of the nominal data. The histogram shows the bootstrap sample, the solid orange line marks the value of $\hat{\alpha}$, and the dashed blue lines mark the 95\% confidence limits.}
   \label{fig:nominal}
\end{figure}

Since the dataset used in this example has missing values, we take this opportunity to explain how the package handles missingness. First, the scores for a given unit of analysis are included in the computation only if two or more scores are present for that unit. Otherwise the unit's row of the data matrix is simply ignored. Second, if two or more scores are present for a given unit, each \texttt{NA} for that unit is ignored in the computations for that row. This is handled both by the loop (adjusted denominator) and by the distance function, which should return 0 if either of its arguments is \texttt{NA}. In the next example we illustrate this by way of a user-defined distance function, and of course the package's built-in distance functions take the same approach.

\subsection{Interval data from an imaging study of hip cartilage}

\begin{figure}[h]
   \centering
   \includegraphics[scale=.3]{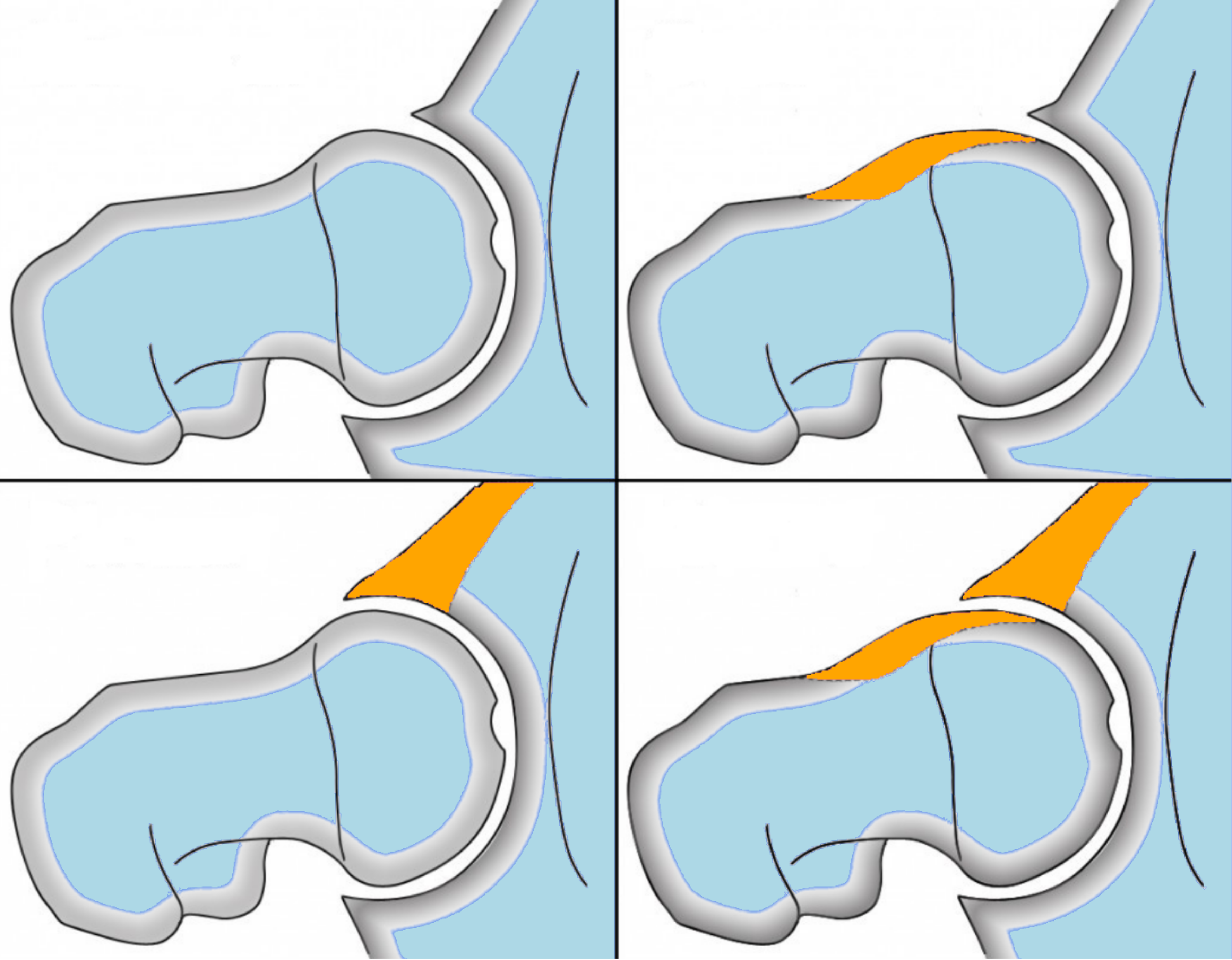}
   \caption{An illustration of femoroacetabular impingement (FAI). Top left: normal hip joint. Top right: cam type FAI (deformed femoral head). Bottom left: pincer type FAI (deformed acetabulum). Bottom right: mixed type (both deformities present).}
   \label{fig:fai}
\end{figure}

The data for this example, some of which appear in Figure~\ref{fig:interval}, are 323 pairs of T2* relaxation times (a magnetic resonance quantity) for femoral cartilage \citep{nissi2015t2} in patients with femoroacetabular impingement (Figure~\ref{fig:fai}), a hip condition that can lead to osteoarthritis. One measurement was taken when a contrast agent was present in the tissue, and the other measurement was taken in the absence of the agent. The aim of the study was to determine whether raw and contrast-enhanced T2* measurements agree closely enough to be interchangeable for the purpose of quantitatively assessing cartilage health.

\begin{figure}[h]
   \centering
   \begin{tabular}{cccccccccccc}
   & $u_1$ &  $u_2$ & $u_3$ & $u_4$ & $u_5$ & $\dots$ & $u_{319}$ & $u_{320}$ & $u_{321}$ & $u_{322}$ & $u_{323}$\vspace{2ex}\\
   $c_1$ & 27.3 & 28.5 & 29.1 & 31.2 & 33.0 & $\dots$ & 19.7 & 21.9 & 17.7 & 22.0 & 19.5\\
   $c_2$ & 27.8 & 25.9 & 19.5 & 27.8 & 26.6 & $\dots$ & 18.3 & 23.1 & 18.0 & 25.7 & 21.7
   \end{tabular}
   \caption{Raw and contrast-enhanced T2* values for femoral cartilage.}
   \label{fig:interval}
\end{figure}

First we load the cartilage data, which are included in the package. The cartilage data are stored in a data frame; we convert the data frame to a matrix, which is the format required by \texttt{krippendorffs.alpha}.

\bigskip\begin{verbatim}
R> data(cartilage)
R> cartilage = as.matrix(cartilage)
\end{verbatim}\bigskip

Now we compute $\hat{\alpha}$ for the interval level of measurement, i.e., squared Euclidean distance. We also produce a bootstrap sample of size 10,000. Since this dataset is much larger than the dataset analyzed in the preceding section, we parallelize the bootstrap computation. We use three CPU cores (of the four available on the author's computer). Setting argument \texttt{verbose} to \texttt{TRUE} causes the fitting function to display a progress bar once again. The computation took five seconds to complete.

\bigskip\begin{verbatim}
R> set.seed(12)
R> fit.sed = krippendorffs.alpha(cartilage, level = "interval", verbose = TRUE,
+                                control = list(bootit = 10000, parallel = TRUE,
+                                nodes = 3))

Control parameter 'type' must be "SOCK", "PVM", "MPI", or "NWS". Setting it to "SOCK".

  |++++++++++++++++++++++++++++++++++++++++++++++++++| 100% elapsed=05s 
\end{verbatim}\bigskip

A call of function \texttt{summary.krippendorffsalpha} produced the output shown below.

\bigskip\begin{verbatim}
R> summary(fit.sed)

Krippendorff's Alpha

Data: 323 units x 2 coders

Call:

krippendorffs.alpha(data = cartilage, level = "interval", verbose = TRUE, 
    control = list(bootit = 10000, parallel = TRUE, nodes = 3))

Control parameters:
              
bootit   10000
parallel TRUE 
nodes    3    
type     SOCK 
              
Results:

      Estimate Lower  Upper
alpha   0.8369 0.808 0.8648
\end{verbatim}\bigskip

We see that $\hat{\alpha}=0.84$ and $\alpha\in(0.81,0.86)$. Thus these data suggest that raw T2* measurements agree almost perfectly with contrast-enhanced T2* measurements, perhaps rendering gadolinium-based contrast agents (GBCAs) unnecessary in T2*-based cartilage assessment. This finding could have clinical significance since the use of GBCAs is not free of risk to patients, especially pregnant women and patients with impaired kidney function. For much additional information regarding the potential risks associated with the use of GBCAs, we refer the interested reader to the University of California, San Francisco's policy on MRI with contrast: \url{https://radiology.ucsf.edu/patient-care/patient-safety/contrast/mri-with-contrast-gadolinium-policy}.

\begin{figure}[ht]
   \centering
   \includegraphics[scale=.65]{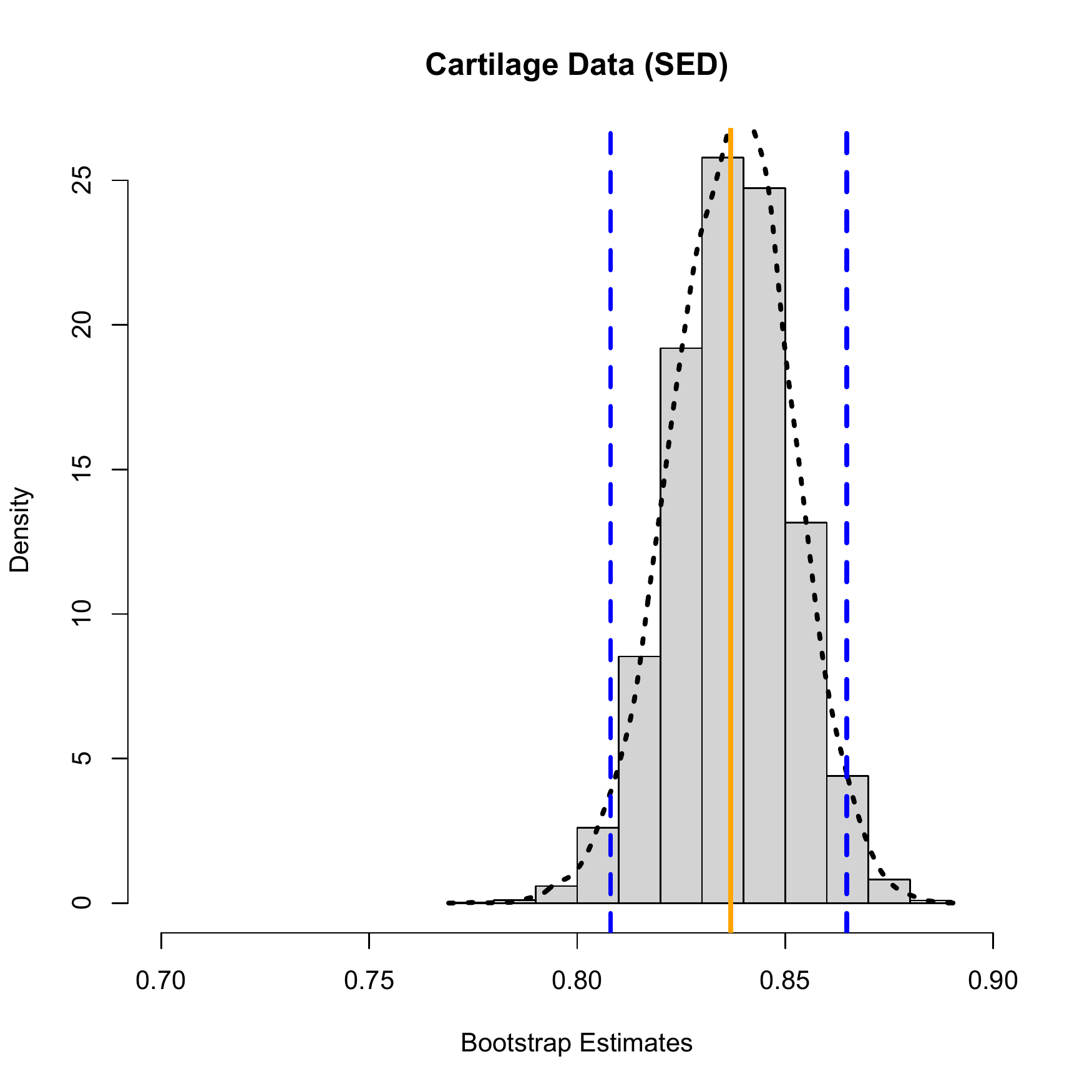}
   \caption{A plot of the results from our analysis of the cartilage data. The histogram and kernel density estimate (dotted black curve) show the bootstrap sample, the solid orange line marks the value of $\hat{\alpha}$, and the dashed blue lines mark the 95\% confidence limits.}
   \label{fig:cartilage}
\end{figure}

Figure~\ref{fig:cartilage} provides a visual display of the cartilage results. The histogram and kernel density estimate show the expected large-sample behavior of $\hat{\alpha}$, i.e., the estimator is approximately Gaussian-distributed and has a small variance.

We mentioned above that attempting to robustify Krippendorff's $\alpha$ by using absolute values in place of squares may prove problematic. This is evident for the cartilage data, as we now demonstrate.

First, define a new distance function as follows. Note that any user-defined distance function must deal explicitly with \texttt{NA}s if the data at hand exhibit missingness. There are no missing values in the cartilage data, but we illustrate the handling of \texttt{NA} anyway.

\bigskip\begin{verbatim}
R> L1.dist = function(x, y)
+ {
+    d = abs(x - y)
+    if (is.na(d))
+        d = 0
+    d
+ }
\end{verbatim}\bigskip

Now we call \texttt{krippendorffs.alpha}, supplying our new distance function via the \texttt{level} argument.

\bigskip\begin{verbatim}
R> fit.L1 = krippendorffs.alpha(cartilage, level = L1.dist, verbose = TRUE,
+                               control = list(bootit = 10000, parallel = TRUE,
+                               nodes = 3))

Control parameter 'type' must be "SOCK", "PVM", "MPI", or "NWS". Setting it to "SOCK".

  |++++++++++++++++++++++++++++++++++++++++++++++++++| 100% elapsed=05s
\end{verbatim}\bigskip

The results are summarized below. These results strongly suggest that only moderate to substantial agreement exists between raw T2* measurements and contrast-enhanced T2* measurements. This contradicts not only our $\alpha_\sed$ analysis but also a Sklar's $\omega$ analysis that assumed a non-central $t$ marginal distribution to accommodate slight asymmetry.

\bigskip\begin{verbatim}
R> summary(fit.L1)

Krippendorff's Alpha

Data: 323 units x 2 coders

Call:

krippendorffs.alpha(data = cartilage, level = L1.dist, verbose = TRUE, 
    control = list(bootit = 10000, parallel = TRUE, nodes = 3))

Control parameters:
              
bootit   10000
parallel TRUE 
nodes    3    
type     SOCK 
              
Results:

      Estimate  Lower Upper
alpha   0.6125 0.5761 0.648
\end{verbatim}\bigskip

\section{Summary and discussion}
\label{summary}

In this article we described Krippendorff's $\alpha$ methodology for measuring agreement, and illustrated the use of R package \texttt{krippendorffsalpha}. We first established $\alpha$'s context among statistical procedures. Specifically, the one-way mixed-effects ANOVA model provides a natural, intuitive genesis for $\alpha$ as the intraclass correlation coefficient. This form of $\alpha$ can be generalized in a parametric fashion to arrive at Sklar's $\omega$, or in a nonparametric fashion to arrive at the form of $\alpha$ presented by Krippendorff, which is a special case of the multiresponse permutation procedure.

We demonstrated the use of \texttt{krippendorffsalpha} version 1.1 by analyzing two datasets: a nominal dataset previously analyzed by Krippendorff, and a sample of raw and contrast-enhanced T2* values from an MRI study of hip cartilage. These analyses highlighted the benefits of the package, which include the use of S3 methods, parallel bootstrap computation, support for user-defined distance functions, and a means of identifying influential units and/or coders.

\section*{Computational details}

The results in this paper were obtained using R~4.0.3 for macOS, and the \texttt{pbapply}~1.4-2 package. R itself and all packages used (save \texttt{kripp.boot}) are available from the Comprehensive R Archive Network (CRAN) at \url{http://CRAN.R-project.org}. Package \texttt{krippendorffsalpha} may be downloaded from CRAN or from the author's GitHub repository, which can be found at \url{https://github.com/drjphughesjr/krippendorffsalpha}. Information about the author's other R packages can be found at \url{http://www.johnhughes.org/software.html}.
  
 \bibliographystyle{apalike}
\bibliography{krippendorffsalpha_Hughes}
  
\end{document}